# SINTERING DILATOMETRY BASED GRAIN GROWTH ASSESSMENT

Charles Manière[a]*, Shirley Chan[a], Geuntak Lee[a,b], Joanna McKittrick[b] and Eugene A. Olevsky[a,c]

[a]*Powder Technology Laboratory, San Diego State University, San Diego, USA;*
[b]*Mechanical and Aerospace Engineering, University of California, San Diego, La Jolla, USA;*
[c]*NanoEngineering, University of California, San Diego, La Jolla, USA*

*corresponding author: CM: Powder Technology Laboratory, San Diego State University, 5500 Campanile Drive, San Diego, CA 92182-1323, , E-mail address: cmaniere@mail.sdsu.edu*

*Microarticle format*

**A B S T R A C T**

Volume shrinkage, grain growth, and their interaction are major events occurring during free sintering of ceramics. A high temperature sintering dilatometry curve is influenced by these both phenomena. It is shown that the continuum theory of sintering can be utilized in the format enabling the extraction of the maximum amount of information on the densification and grain growth kinetics based on a simple dilatometry test. We present here the capability of such a fast approach (Dilatometry based Grain growth Assessment DGA) utilized for the modeling of sintering and grain growth of zirconia.

## I. Introduction

During free sintering of ceramics the grain growth usually occurs at the final stage of processing when relative density is close to 90 %. When grain growth starts, it affects the densification kinetics because the grain size has a direct influence on the length of the diffusional path and on the capillary forces that govern sintering. In consequence, the grain growth slows down the densification of the powder specimen at the final stage of sintering. Conversely, the sintering has also an influence on the grain growth kinetics given that the porosity or the segregation of impurities may reduce the grain growth rate [1,2]. Previously it was shown that taking into account the grain growth improves the final density predictions by a sintering model describing spark plasma sintering experiments [3,4]. Accurate assessment of the grain growth, highly active at the end of sintering, is of great importance because grain growth favors residual porosity and may reduce the mechanical properties of the processed material. In other advanced sintering approaches, controlling the grain growth/densification behavior (also called sintering trajectory) is the key issue. These approaches include the two-step sintering method which prescribes the optimized processing thermal history to reduce the final grain size, the pressure and/or field assisted sintering, such as spark plasma sintering, which highly improves the retention of the small grain size microstructures [5,6]. The modeling of both densification and grain growth is therefore of high interest for these techniques. In the present study, the full extraction of the densification and grain growth constitutive parameters is conducted based on a simple free sintering dilatometry curve for zirconia. Such an analysis is made possible due to the developed specific formulation based on the continuum theory of sintering [7] adopted for the description of the material constitutive behavior at the intermediate and final stages of sintering.

## II. Theory and method

Free sintering can be modelled analytically via the following equation of the continuum theory of sintering [7]:

$$P_l = \frac{-A(T)\psi\dot{\theta}}{(1-\theta)} = \frac{3\alpha(1-\theta)^2}{r} \qquad (1)$$

where $P_l$ is the sintering stress whose expression is detailed in the third term, $A(T)$ is the material viscosity, $\psi$ is the bulk modulus, $\theta$ is the porosity, $\alpha$ is the surface energy and $r$ is the grain radius.

This equation can be reformulated to extract the temperature dependent viscosity $A(T)$ and the surface energy $\alpha$ for the intermediate sintering stage using the regression equation:

$$Y = ln\left(\frac{-3(1-\theta)^3}{rT\psi\dot\theta}\right) = ln\left(\frac{A(T)}{\alpha}\right) = ln\left(\frac{A_0}{\alpha}\right) + \frac{Q}{RT} \qquad (2)$$

where $R$ is the gas constant, $T$ is the absolute temperature, $A_0$ and $Q$ are the Arrhenius pre-exponential viscosity constant and activation energy, respectively.

At the final sintering stage, when the grain growth occurs, the interaction between the sintering stress $P_l$ and grain growth slowdowns the sintering kinetics [4]. Based on this assumption and using a specific formulation of the continuum theory of sintering, it is possible to estimate the grain size evolution during the final stage of sintering via the middle term of the following equation (knowing $\frac{A_0}{\alpha}$ and $Q$ from the intermediate stage and the bulk modulus $\psi = \frac{2(1-\theta)^3}{3\theta}$ from ref [7]):

$$G = \frac{-6(1-\theta)^3}{\frac{A_0}{\alpha}exp\left(\frac{Q}{RT}\right)T\psi\dot\theta} = \left(\frac{k_0 exp\left(\frac{-Q_G}{RT}\right)}{\dot G}\right)^{1/3} \qquad (3)$$

where, $G$ is the grain diameter, $k_0$ and $Q_G$ are the grain growth Arrhenius preexponential constant and activation energy, respectively. The last member of equation (3) is the grain growth kinetic term depending on its unknown parameters ($k_0$ and $Q_G$). Knowing the estimated grain size at the final stage of sintering, it is finally possible to determine the grain growth kinetic parameters using the new regression equation:

$$Y_G = ln\left(\left(\frac{-6(1-\theta)^3}{\frac{A_0}{\alpha}exp\left(\frac{Q}{RT}\right)T\psi\dot\theta}\right)^3 \dot G\right) = ln(k_0) - \frac{Q_G}{RT} \qquad (4)$$

### III. Experimental procedure

The sintering in air of Tosoh zirconia (TZ-3YS-E, 50 nm) powder has been carried out in the dilatometer (Unitherm model 1161, Anter Corporation). The sintering curves have been calculated using the sample's shrinkage and tooling thermal expansion curves (see Figure 1a). A 5 K/min ramping cycle up to 1700 °C is chosen to ensure the sintering up to high density and with the pronounced grain growth. The sample has been prepared by cold isostatic pressing at 400 MPa without binder to ensure a good initial packing of the powder.

### IV. Results and discussion

The densification curve reported in Figure 1a shows, as expected, the sintering slowdown that appears in the 87-95 % (relative density) region. Assuming that the densification deceleration, that happens earlier than the full density is achieved, is caused by the grain growth interaction with the densification, the sintering constitutive parameters can be determined using the following step by step procedure. At the first stage, the densification parameters are determined through the regression equation (2) for the kinetic curve region not influenced by the grain growth (between 55 and 87% of the relative density). The regression graph Y *vs* 1/(RT) reported in Figure 1b gives values of $Ao$ = 22.67 Pa m$^2$ s K$^{-1}$ J$^{-1}$ and $Q$ = 220.1 kJ/mol. The resulting densification model without grain growth (reported in Figure 1c) explains well the densification behavior at the intermediate stage of sintering, but the sintering slowdown during the final stage is not reproduced. The strategy of equation (3) is to estimate the grain growth accounting for the difference between the relative density at the sintering final stage predicted by the classical model and the experimentally obtained relative density value (which is lower). This gives the graph of the estimated G *vs* T (Figure.1b) via equation (3). Finally, knowing $G$ and $\dot G$, it is possible to determine the grain growth parameters ($k_0$ and $Q_G$) using the regression equation (4). The regression graph Y$_G$ *vs* 1/(RT) (reported in Figure 1b) exhibits two trends of the grain growth behavior. In the temperature range of 1700-1850K (zone 1) we obtained $k_0$ = 1.100E+25 m$^4$/s and $Q_G$ = 1879 kJ/mol, while in the temperature range of 1850-1973K (zone 2) these values evolve to $k_0$ = 2.741E-14 m$^4$/s and $Q_G$ = 514.4 kJ/mol. The values of $Q_G$ in zone 2 are close to the grain growth activation energy earlier reported (520 kJ/mol [8], 584 kJ/mol [9]). This seems to show the two mechanisms of the grain growth: in zone 1 the grain growth is disturbed by the high level of porosity and shows the higher activation energy than the second mechanism in zone 2 where the activation energy matches the literature data [8,9].

Finally, the new sintering constitutive model (Figure 1c) describes well the densification during the initial, intermediate and final stages of sintering. The modeling of the grain growth reported in Figure 1d shows a good agreement between the two calculations approaches: one based on equations (1) and (3) and the other one based on the combination of equations (1) and (4).

**IV. Conclusions**

The developed approach shows that the identification of the major sintering and grain growth constitutive parameters can be conducted concurrently using a simple dilatometry curve. This becomes possible due to an optimized analysis of the intermediate (densification only) and final (densification and grain growth) sintering kinetics. The sintering/grain growth modeling successfully predicts the significant reduction of the densification rate during the final stage of sintering due to the grain growth. The grain growth exhibits two distinct stages: a transient stage for the high level of porosity, and the final stage (that appears after 91 % of the relative density is reached) at which the grain growth impacts the regular sintering kinetics. It should be noted that the developed method identifies the grain growth kinetics specifically during the final stage of sintering; therefore, the determined constitutive parameters are restricted to this small experimental domain. Nevertheless, the simplicity of this approach makes it useful for the refinement of the densification model for the final stage of sintering where the densification slowdown due to the grain growth is very common. This method can be also applied to obtain the grain growth kinetics from the dilatometry data.

**Acknowledgements**

The support of the US Department of Energy, Materials Sciences Division, under Award No. DE-SC0008581 is gratefully acknowledged.

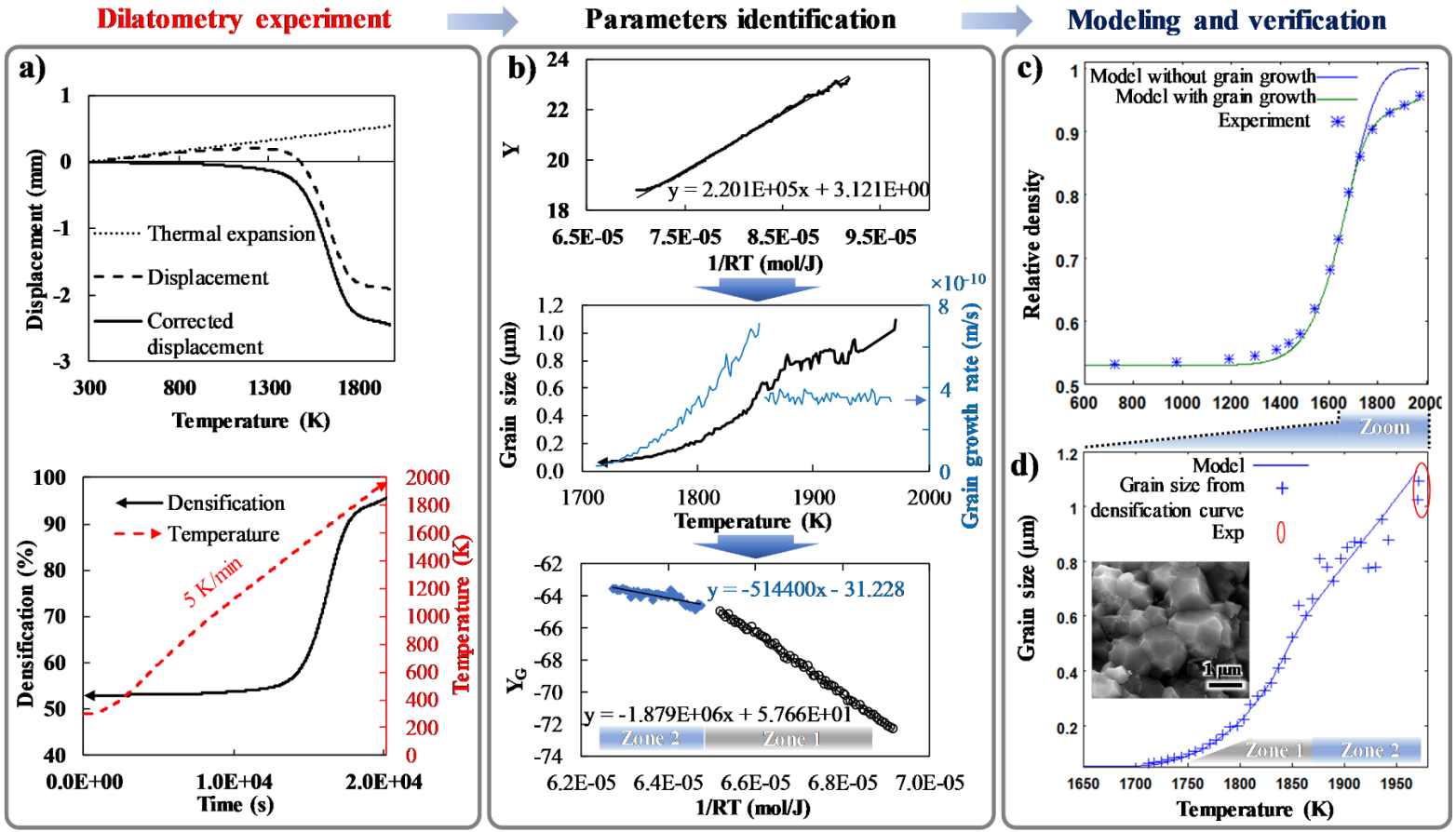

Figure 1. Sintering/grain growth parameters identification, a) dilatometry experiment, b) determination of the densification and grain growth parameters, c) modeling and verification of the densification and d) grain size evolution curves.